\documentclass[11pt]{article} 
\usepackage{amssymb}
\usepackage{epsfig}
\newlength{\bredde} 
\def\slash#1{\settowidth{\bredde}{$#1$}\ifmmode\,\raisebox{.15ex}{/}  
\hspace*{-\bredde} #1\else$\,\raisebox{.15ex}{/}\hspace*{-\bredde} #1$\fi}  
\newcommand{\sect}[1]{\setcounter{equation}{0}\section{#1}}

\begin{document}  
\topmargin -1.4cm  
\oddsidemargin -0.8cm  
\evensidemargin -0.8cm  
\title{\Large{{\bf   
Prediction of RNA pseudoknots by Monte Carlo simulations  
}}}  
  
\vspace{1.5cm}  
\author{~\\{\sc G. Vernizzi}$^1$, {\sc H. Orland}$^1$  
and {\sc A. Zee}$^2$\\~\\  
$^1$Service de Physique Th\'eorique, CEA/DSM/SPhT Saclay\\  
Unit\'e de recherche associ\'ee au CNRS\\  
F-91191 Gif-sur-Yvette Cedex, France\\~\\  
$^2$Institute of Theoretical Physics and Department of Physics\\  
University of California, Santa Barbara, CA 93106, USA  
}

\date{}  
\maketitle  
\vfill  
\begin{abstract}  
In this paper we consider the problem of RNA folding with pseudoknots.
We use a graphical representation in which the secondary structures
are described by planar diagrams. Pseudoknots are identified as
non-planar diagrams. We analyze the non-planar topologies of RNA
structures and propose a classification of RNA pseudoknots according
to the minimal genus of the surface on which the RNA structure can be
embedded.  This classification provides a simple and natural way to
tackle the problem of RNA folding prediction in presence of
pseudoknots. Based on that approach, we describe a Monte Carlo
algorithm for the prediction of pseudoknots in an RNA molecule.
\end{abstract}  


\vfill

\begin{flushleft}  
SPhT-T04/061\\ 
q-bio.BM/0405014  
\end{flushleft}  
\thispagestyle{empty}  
\newpage

\renewcommand{\thefootnote}{\arabic{footnote}}  
\setcounter{footnote}{0}

\sect{Introduction}
\label{introsection}  
In recent years the quest for an algorithm which can predict the
spatial structure of an RNA molecule given its chemical sequence has
received considerable attention from molecular biologists
\cite{Science}. In fact the three-dimensional structure of an RNA
molecule is intimately connected to its specific biological function
in the cell (e.g. for protein synthesis and transport, catalysis,
chromosome replication and regulation) \cite{TB}. It is determined by
the sequence of nucleotides along the sugar-phosphate backbone of the
RNA. The chemical formula or sequence of covalently linked nucleotides
along the molecule from the 5' to the 3' end is called the {\it
primary structure}. The four basic types of nucleotides are adenine
(A), cytosine (C), guanine (G) and uracil (U), but it is known that
modified bases may appear \cite{LCC}.
 
At high enough temperatures, or under high-denaturant conditions RNA
molecules have the three-dimensional structure of a free
single-stranded swollen polymer. At room temperature, different
nucleotides can pair by means of saturating hydrogen bonds. The
standard Watson-Crick pairs are A$\bullet$U and C$\bullet$G with two
and three hydrogen bonds respectively, whereas G$\bullet$U is a wobble
pair with two hydrogen bonds. Comparative methods showed that
``non-canonical'' pairings are also possible \cite{Nagaswamy}, as well
as higher-order interactions such as triplets, or quartets.  In this
paper we will consider only canonical base-pair interactions. Adjacent
base pairs can stack, providing and additional binding energy which is
actually the origin of the formation of stable A-form helices, one of
the main structural characteristics of folded RNAs. Helices may embed
unpaired sections of RNA, in the form of hairpins, loops and bulges.
It is all these pairings, stackings of bases and structural motifs
which bring the RNA into its folded three-dimensional
configuration. One of the main open problems of molecular biology is
the prediction of the actual spatial molecular structure of RNA
(i.e. its {\it shape}) given its primary structure.

As we shall see in Section \ref{representsection}, it is possible to
define {\it secondary structures} of RNA as structures in which the
pairings between canonical base pairs do not cross in a certain
representation (planar graphs).  One can also define the {\it tertiary
structure} of RNA which is the actual three-dimensional arrangement of
the base sequence.  This classification corresponds to the fact that
the secondary structure of RNA carries the main contribution to the
free energy of a fully folded RNA configuration, including also some
of the sterical constraints. For that reason one can attempt to
describe the folding process hierarchically \cite{TB,BJT,BW,LD}.
However, since the secondary structure describes just the topology of
binary contacts of the bases, most of the information about distances
in real three-dimensional space is lost.  The importance of the
secondary structure relies in the fact that it may provide the
``skeleton'' of the final tertiary structure.
 
Over the past twenty years several algorithms have been proposed for
the prediction of RNA folding.  They are based on: deterministic or
stochastic minimization of a free energy function \cite{ZS,Monte},
phylogenetic comparison \cite{phyl1,phyl2,phyl3,phyl4}, kinetic
folding \cite{FFHS,mironovD,isa1,isa2}, maximal weighted matching
method \cite{MWM}, and several others (for a survey see \cite{Z2}).
It is fair to say that despite the large number of tools available for
the prediction of RNA structures, no reliable algorithms exist for the
prediction of the full tertiary structure of RNA. Most of the
algorithms listed above deal with the prediction of just the RNA
secondary structure.  To describe the full folding it is important to
introduce the concept of RNA pseudoknot \cite{PRB}. One says that two
base pairs form a pseudoknot when the parts of the RNA sequence
spanned by those two base pairs are neither disjoint, nor have one
contained in the other.  Thus RNA secondary structures without
pseudoknots can be represented by planar diagrams, whereas RNA with
pseudoknots appear when two base pairs can ``cross'', leading to
non-planar diagrams (a more precise definition is given in the next
Section). Pseudoknots play an important role in natural RNAs
\cite{WJ}, for structural, regulatory and catalytic
functions. Pseudoknots are excluded in the definition of RNA secondary
structure and many authors consider them as part of the tertiary
structure. This restriction is due to the fact that RNA secondary
structures without pseudoknots can be predicted easily. One should
also note that pseudoknots very often involve base-pairing from
distant parts of the RNA, and are thus quite sensitive to the ionic
strength of the solution. It has been shown that the number of
pseudoknots depends on the concentration of Mg$^{++}$ ion, and can be
strongly suppressed by decreasing the ionic strength (thus enhancing
electrostatic repulsion) \cite{MD,Dave,MD2}.  The most popular and
successful technique for predicting secondary structures is dynamic
programming \cite{ZS,waterman,williams,NJ,Z,Z1,HH2,Wuchty}, for which
the memory and CPU requirements scale with the sequence length $L$ as
$O(L^2)$ and $O(L^3)$, respectively.
 
Recently, new deterministic algorithms that deal with pseudoknots have
been formulated
\cite{RE,Uemura,Akutsu,LP,giege,deogun,OZ,POZ,PTOZ}. In this case the
memory and CPU requirements generally scale as $O(L^4)$ and $O(L^6)$
respectively ($O(L^4)$ and $O(L^5)$ in
\cite{LP}, or $O(L^4)$ and $O(L^3)$ for a restricted model in
\cite{Akutsu}), which can be a very demanding computational effort
even for short RNA sequences ($L\sim100$). Moreover, the main
limitation of these algorithms is the lack of precise experimental
informations about the contribution of pseudoknots to the RNA free
energy, which is often excluded a priori in the data analysis (as also
pointed out in \cite{isa1,MironovL,gultyaev}).  The
increase of computational complexity does not come as a surprise. In
fact the RNA-folding problem with pseudoknots has been proven to be
NP-complete for some classes of pseudoknots \cite{Akutsu,LP}.
For that reason, stochastic algorithms might be a better choice to
predict secondary structures with pseudoknots in a reasonable time and
for long enough sequences.

In \cite{Monte,abrah,gultyMonte,Ivo} stochastic Monte Carlo algorithms
for the prediction of RNA pseudoknots have been proposed.  In these
stochastic approaches, the very irregular structure of the energy
landscape (glassy-like) is the main obstacle: configurations with
small differences in energy may be separated by high energy barriers,
and the system may very easily get trapped in metastable states. Among
the stochastic methods, the direct simulation of the RNA-folding
dynamics (including pseudoknots) with kinetic folding algorithms
\cite{isa1,isa2} is most successful. This technique allows to
describe the succession of secondary structures with pseudoknots
during the folding process. The approach we follow in this paper is
close in spirit to that one, with a stronger emphasis on the
topological character of the RNA pseudoknots.  It is based on a
correspondence (first noticed by E. Rivas and S.R. Eddy in \cite{RE})
between a graphical representation of RNA secondary structures with
pseudoknots and Feynman diagrams.  In \cite{RE} the authors consider
only a particular class of pseudoknots.  Along the same direction, the
authors of \cite{OZ} made the correspondence between RNA folding and
Feynman diagrams more explicit by formulating a {\it matrix field
theory} model whose Feynman diagrams give exactly all the RNA
secondary structures with pseudoknots. The remarkable facts of this
new approach is that it provides an analytic tool for the prediction
of pseudoknots, and all the diagrams appear to be naturally organized
in a series of terms, called the {\it topological expansion}, where
the first term corresponds to planar secondary structures without
pseudoknots, and higher-order terms correspond to structures with
pseudoknots.
 
In this paper we explore in more detail this topological expansion
and its potential predictive power.  We also propose a numerical
stochastic algorithm for dealing with this expansion in a systematic
way, which in principle allows the prediction of all kinds of RNA
pseudoknots. The paper is organized as follows.  In Section
\ref{representsection} we review some well-known graphical
representation of RNA structures, with special emphasis on the
so-called ``disk diagram'' representation. In such a representation
one can uniquely associate to each RNA secondary structure with (or
without) pseudoknots, a circle diagram which is planar (or not planar, 
respectively).  In Section \ref{toposection} we show how one can
characterize the ``degree of non-planarity'' of a given disk diagram.
In fact, one can always associate an integer number to each RNA disk
diagram, called the {\it genus}, and we will describe its topological
meaning and information content. We thus propose to classify
RNA pseudoknots according to their genus.  Following this idea, in Section
\ref{statsection} we generalize the standard thermodynamic model for the
description of RNA structures to the inclusion of pseudoknots. The
generalized model we propose is very natural, in the same spirit when
going from the Canonical Ensemble to the Grand Canonical Ensemble in
statistical mechanics. Our model can control the topological
fluctuations i.e. the formation of pseudoknots in the RNA molecule,
and we will describe the general features of its phase diagram.  In
Section \ref{montesection} we describe a Monte Carlo algorithm for the
actual calculation of thermodynamical quantities in our generalized
model. In particular we will list in details the Monte Carlo moves,
the free-energy updating algorithm and the simulated annealing method
we propose for dealing with the problem of high energy
barriers. Section \ref{conclsection} contains the concluding remarks,
and the Appendix is devoted to the explicit description of a part of
the Monte Carlo algorithm of Section
\ref{montesection}.

\sect{Representation of RNA structures}  
\label{representsection} 
 
Any RNA sequence can be represented as the list of nucleotides
$r_i\in$(A,C,G,U), $i=1,\ldots,L$, where $r_i$ is the $i$-th
nucleotide along the oriented sugar backbone (from 5' to 3'). The
ordered list $\{r_1,r_2,\ldots,r_L\}$ is called the primary structure
of the RNA.
  
\begin{figure}[-t]  
\centerline{  
\epsfig{figure=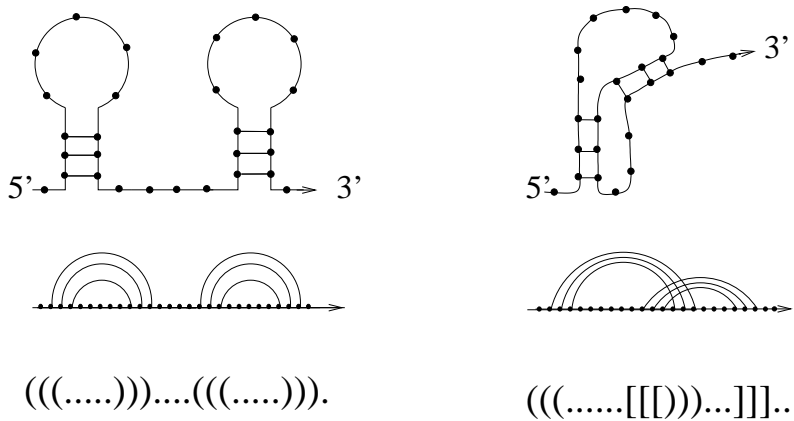,width=25pc}  
}  
\caption{ 
An RNA configuration without pseudoknots (left column) and RNA
configuration with a simple ``H'' pseudoknot (right column).  From the
top to the bottom: the RNA configuration, arc representation and
bracket representation. Note that the arc diagram for pseudoknotted
RNA has crossing arcs and the bracket representation requires two
kinds of parenthesis.}
\label{bracketplot}  
\end{figure}

The RNA secondary structure requires a more graphical representation.
Actually there are several equivalent ways to represent an RNA
secondary structure with a given primary structure. The most commonly
used representation is perhaps the {\it bracket} notation, where two
paired bases, $r_i$ and $r_j$ ($i<j$), are represented by parenthesis
``(`` and ``)'', and unpaired bases are represented by a dot '.' or a
colon ':' (see Figure
\ref{bracketplot}). Pseudoknots can be described in a similar fashion, 
but one needs to introduce several different kinds of brackets (like
square brackets '[', ']' or braces '$\{$', '$\}$', as for example in
the database \cite{pseudobase,pseudobase1}), and this is not a
very efficient representation for complicated structures.

\begin{figure}[-t]  
\centerline{  
\epsfig{figure=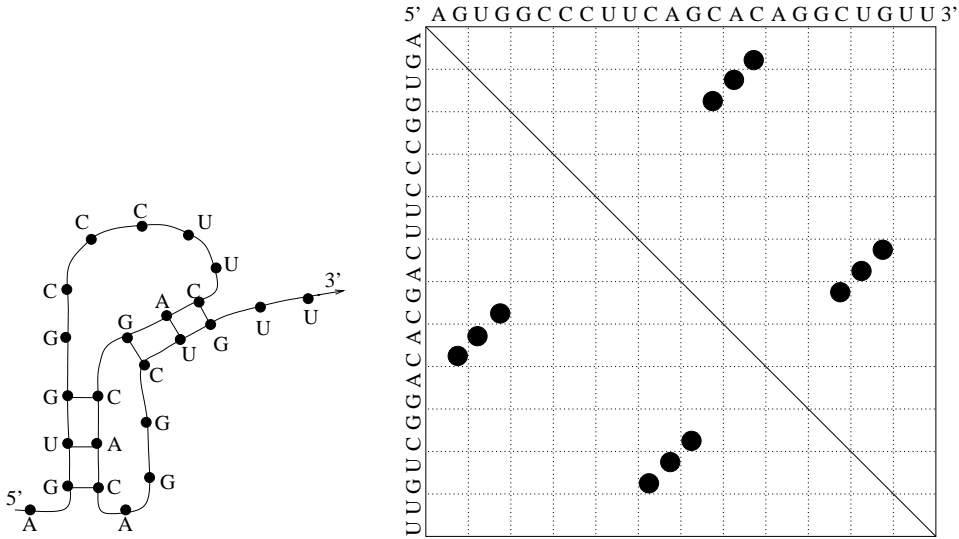,width=30pc}  
}  
\caption{Representation of an RNA secondary structure with an ``H'' pseudoknot (left), and the corresponding dot plot diagram (right).} 
\label{rnadotplot}  
\end{figure}

Among several other representations (e.g. mountain diagrams
\cite{HH}, tree diagrams \cite{FKSS}, graphs \cite{RNAgraph1}), 
a very general and widely used representation is the so-called {\it
dot plot} diagram. It is an array where a dot is placed in the row $i$
and column $j$ if the bases $r_i$ and $r_j$ are actually paired (see
Figure
\ref{rnadotplot}). This plot is the graphical representation of the $L
\times L$ {\it contact matrix} $C$ with elements
\begin{equation} 
C_{ij}= 
\left\{ 
\begin{array}{ll} 
1 & \mbox{if } i \mbox{ and } j  \mbox{ are paired} \, ,\\ 
0 & \mbox{ otherwise.} 
\end{array} 
\right. 
\label{Contact} 
\end{equation}  
In mathematical terms, the contact matrix $C$ is the matrix 
of the permutation involution associated to the 
given set of pairings. In fact, one can always interpret the base pairing 
$i-j$ as a transposition of the elements $\{i,j\}$ and therefore one 
can uniquely associate a permutation $\sigma$ to any structure by: 
\begin{equation} 
\sigma(i)= 
\left\{ 
\begin{array}{ll} 
j & \mbox{if } i \mbox{ and } j  \mbox{ are paired} \, ,\\ 
i & \mbox{ otherwise.} 
\end{array} 
\right. 
\end{equation}  
For example, if the primary structure is $\{5'-CUUCAUCAGGAAAUGAC-3' 
\}$ and the pseudoknotted secondary structure is: $.(((.[[[)))..]]].$, 
one can associate to it the permutation: 
\begin{equation} 
\sigma=\left(  
\begin{array}{ccccccccccccccccc}  
.&(&(&(&.&[&[&[&)&)&)&.&.&]&]&]&.\\  
1& 2 &3 &4 &5 &6 &7 &8 &9 &10& 11& 12& 13& 14& 15& 16& 17  \\  
1&11 &10&9&5 &16&15&14 &4 &3 &2  & 12& 13& 8 & 7 & 6 & 17   
\end{array} 
\right) \, , 
\end{equation}  
which is also an involution since $\sigma^2$ is the identity 
permutation.  The matrix representation of $\sigma$ is the matrix $D$ 
with $D_{i, \sigma(i)}=1$ and $0$ otherwise. Obviously $D=C+{\cal I}$, 
where ${\cal I}$ is the $L \times L$ identity matrix. This notation is 
very useful for numerical implementations of the algorithm we propose 
in Section \ref{montesection}. The advantage of the dot plot diagram 
is that it allows the comparison between different RNA secondary 
structures, just by superimposition as it is necessary for 
comparative analysis. Moreover it can be used for representing 
RNA structure with any kind of pseudoknots. 
  
\begin{figure}[-ht]  
\centerline{  
\epsfig{figure=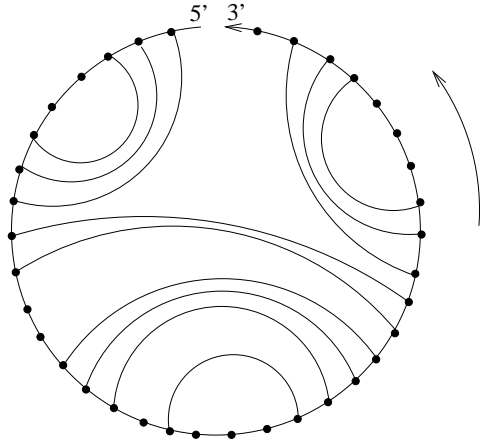,width=15pc}  
}  
\caption{ 
Typical disk (circle) diagram representation of an RNA secondary 
structure without pseudoknots. The circle is anticlockwise oriented 
from $5'$ to $3'$.  Note that there are no crossing arcs.} 
\label{circleplot}  
\end{figure}

\begin{figure}[-ht]  
\centerline{  
\epsfig{figure=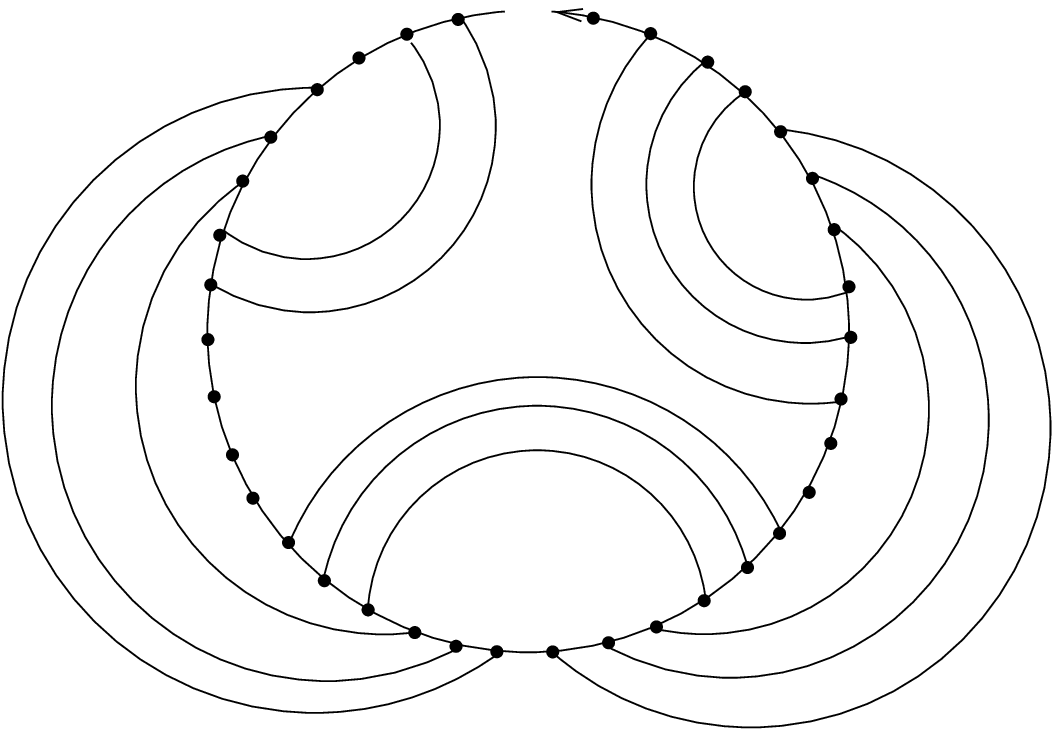,width=11pc}  
\epsfig{figure=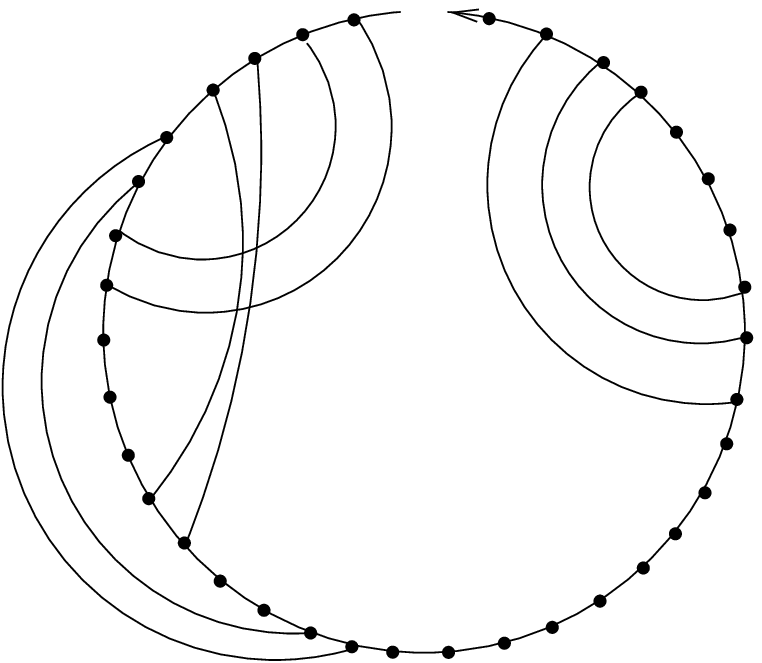,width=9pc}  
\epsfig{figure=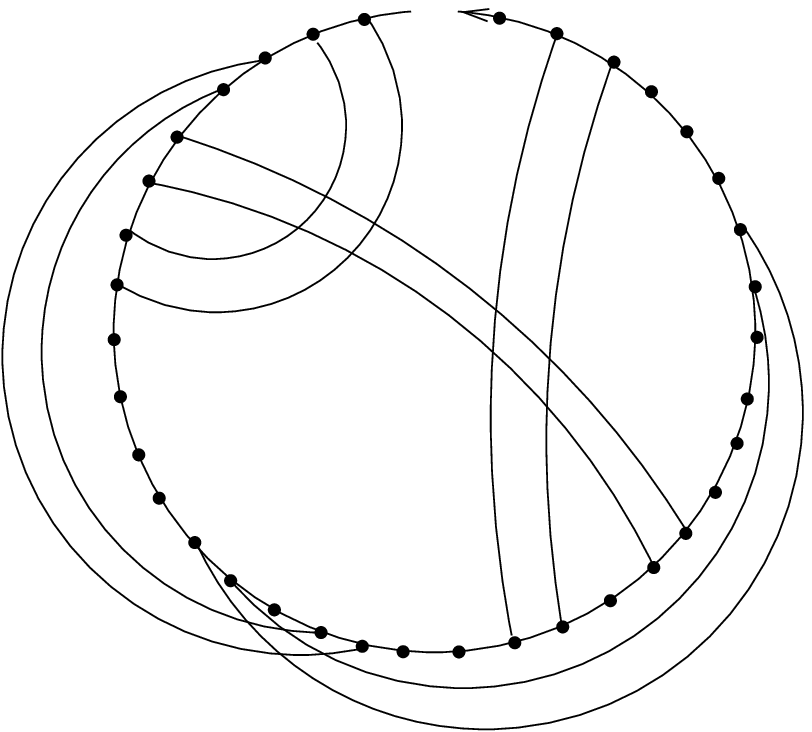,width=9pc}  
\put(-290,-13){$a)$} 
\put(-170,-13){$b)$} 
\put(-55,-13){$c)$}  
} 
\caption{ 
Three kind of disk diagrams for RNA secondary structures with
pseudoknots. The authors of \cite{RE} consider cases of the form a)
(``overlapping pseudoknots''), b) (pseudoknot present in {\it
Escherichia coli $\alpha$} mRNA \cite{Gluick}) but not c) (parallel
$\beta$-sheet protein interaction). The technique in
\cite{OZ} can deal with all the three cases.}
\label{rivaseddyplot}  
\end{figure}

A representation which is completely equivalent to the dot plot
diagram is the {\it disk diagram} (also called {\it circle plot} or
{\it circular plot}). In this case the RNA sequence is represented as
an oriented circle (from 5' to 3') by virtually linking the first
nucleotide to the last one. Each base pairing is represented as an arc
inside the circle, connecting the two paired bases. Figure
\ref{circleplot} shows a typical disk diagram.  In this representation
secondary structures without pseudoknots are purely planar diagrams,
i.e. diagrams that can be drawn without crossing arcs, whereas
pseudoknots correspond to structures which are not planar.
This fact has been already observed by E.Rivas and S.R.Eddy in 
\cite{RE}, where they consider diagrams with arcs inside 
{\it and} outside the disk\footnote{More precisely, they represent the
RNA sequence as an oriented straight line, and the pairings as arcs
above and below that line. This is of course equivalent to the disk
representation.}. Crossing arcs are allowed but only inside or only
outside but not both at the same time (so-called ``overlapping
pseudoknots'', see diagrams a) and b) of Figure
\ref{rivaseddyplot}). As it was shown in
\cite{POZ,PTOZ}, several general classes of pseudoknots cannot be  
described in such a simple way (such as the diagram on the right of Figure  
\ref{rivaseddyplot}). It is then more convenient to draw the 
arcs always inside the disk (or outside, but not both) and to consider 
all the corresponding diagrams as non planar.  It is precisely 
following this approach that the authors of \cite{OZ} found an 
algorithm for computing pseudoknots with matrix field theory in a 
completely general fashion.  In this paper we pursue the same analysis 
by considering the diagrams themselves and not the associated matrix 
field theory model.\\ 

\sect{The topological character of RNA pseudoknots}
\label{toposection} 
  
There is a very natural way for classifying the ``degree of  
non-planarity'' of a given disk diagram, which we review here  
briefly. It is based on a topological analysis introduced long ago by  
Euler. We emphasize that this characterization is a well-known  
classical result of algebraic topology  
and it has been already introduced in \cite{OZ} for RNA secondary  
structures. We repeat it here in more detail for the convenience  
of the reader.  
  
\begin{figure}[-t]  
\centerline{  
\epsfig{figure=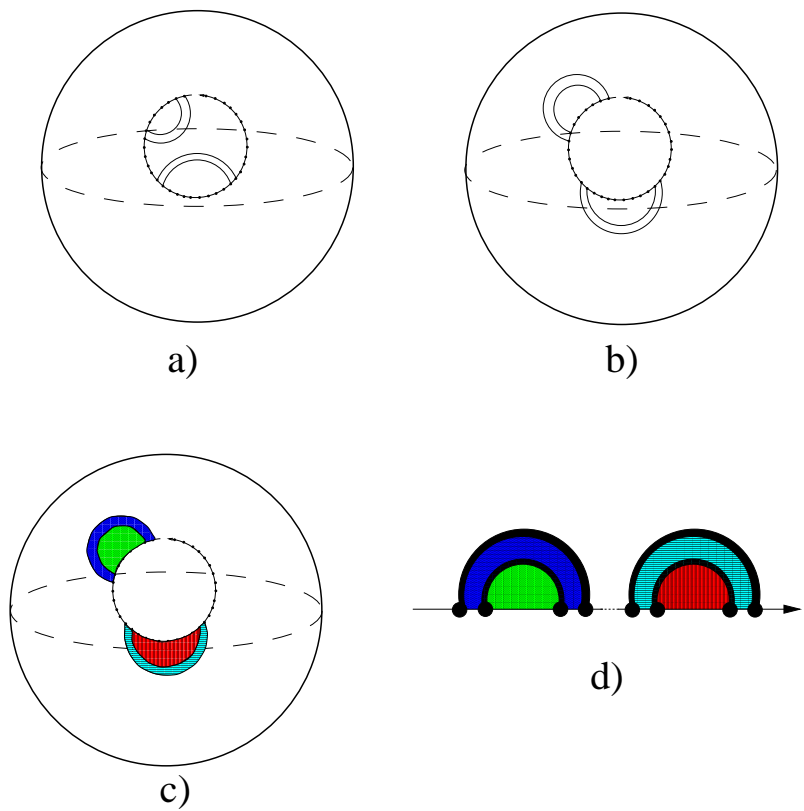,width=25pc}  
}  
\caption{Example of planar disk diagram. In a) the disk diagram of a double hairpin (like the one in Figure \ref{bracketplot}) is on a sphere. In b) the arcs are drawn all outside the circle. In c) the sphere is partitioned in 6 patches (5 faces and one ``hole'', i.e. the RNA circle). In d) is the representation of c) in double line notation (black thick arcs). Here $\#F=5$, $\#V=4$, $\#E=8$, and therefore $\chi=1$, i.e. $g=0$.} 
\label{sphereplot}  
\end{figure}

As we have shown, in any disk diagram the RNA sequence is represented  
by an oriented circle. When the circle is drawn on a sphere, its  
orientation allows to distinguish an ``inside'' and an ``outside'' of  
the circle.  One says that the circle is a ``boundary'' or  
``puncture'' (as it can be drawn smaller and smaller, in a continuous  
fashion up to single point) on the sphere.  Hence any (disk) planar  
diagram can be drawn on a sphere without crossing lines, simply by  
drawing the arcs on the same side (see Figure \ref{sphereplot}). The key  
observation is that the sphere is naturally partitioned in several  
parts by the diagram. As explained in \cite{OZ} it is useful to draw  
the arcs with a ``double-line notation'' (see Figure \ref{sphereplot}). In  
this way it is clear that the sphere is partitioned into several  
polygons. Note that all the lines have an orientation induced by the  
one of the circle.  
  
The Euler characteristic $\chi$ of a  diagram is defined as   
\begin{equation}  
\chi=\#V-\#E+\#F \, ,  
\label{genus}  
\end{equation}  
where $\#V$, $\#E$ and $\#F$ are the numbers of vertices,  
edges, and faces, respectively. A vertex is just a nucleotide, an edge is any line  
connecting two nucleotides (either an arc joining the nucleotides, or  
the RNA sequence) and a face is that part of the surface within a closed  
loop of edges. Obviously, if there are $n$ arcs then $\#E=\#V+n$.  A  
famous theorem of Euler states that any polyhedron homeomorphic to a  
sphere with a boundary (puncture) has an Euler characteristic  
$\chi=1$. Therefore all RNA secondary structures without pseudoknots  
are described by disk diagrams with $\chi=1$.

\begin{figure}[-ht]  
\centerline{  
\epsfig{figure=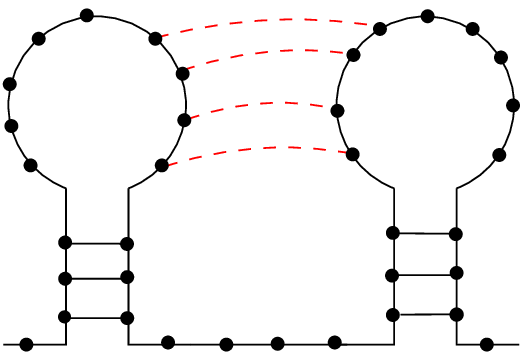,width=12pc}  
\epsfig{figure=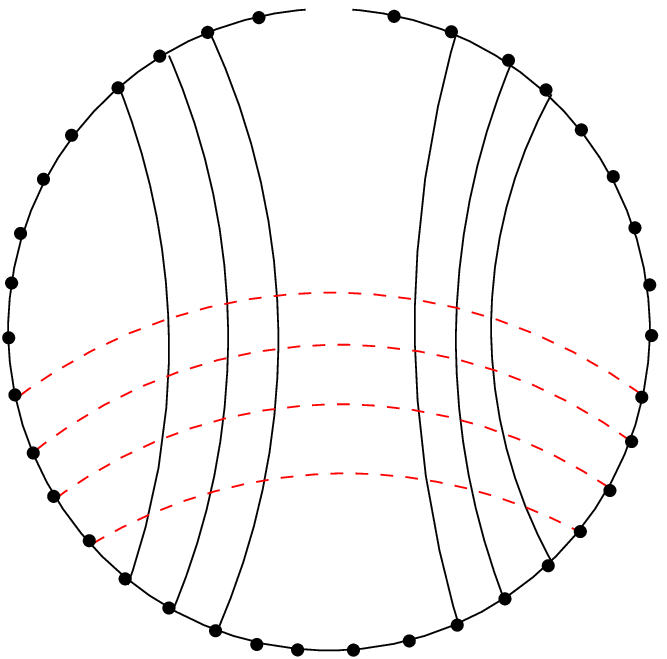,width=10pc}  
}  
\caption{A ``kissing hairpin'' pseudoknot. The respective disk diagram has 
crossing arcs necessarily, when the arcs are drawn all inside (or all 
outside) the RNA circle.} 
\label{kisshairpinplot}  
\end{figure}
\begin{figure}[-ht]  
\centerline{  
\epsfig{figure=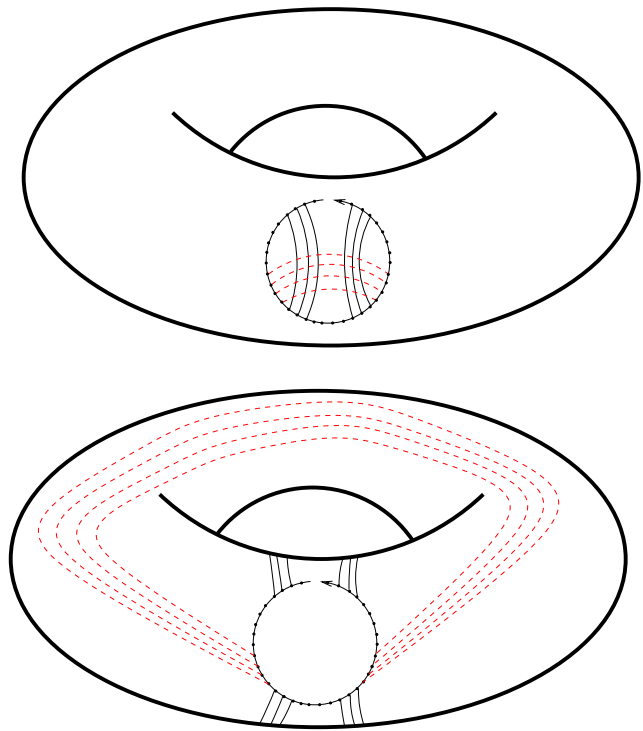,width=20pc}  
}  
\caption{The ``kissing hairpins'' of Figure \ref{kisshairpinplot} can be drawn on a torus without intersections. In this example $\#F=9$, $\#V=20$, $\#E=20+10$, and therefore $\chi=-1$, i.e. $g=1$.}  
\label{torusplot}  
\end{figure}

\begin{figure}[-ht]  
\centerline{  
\epsfig{figure=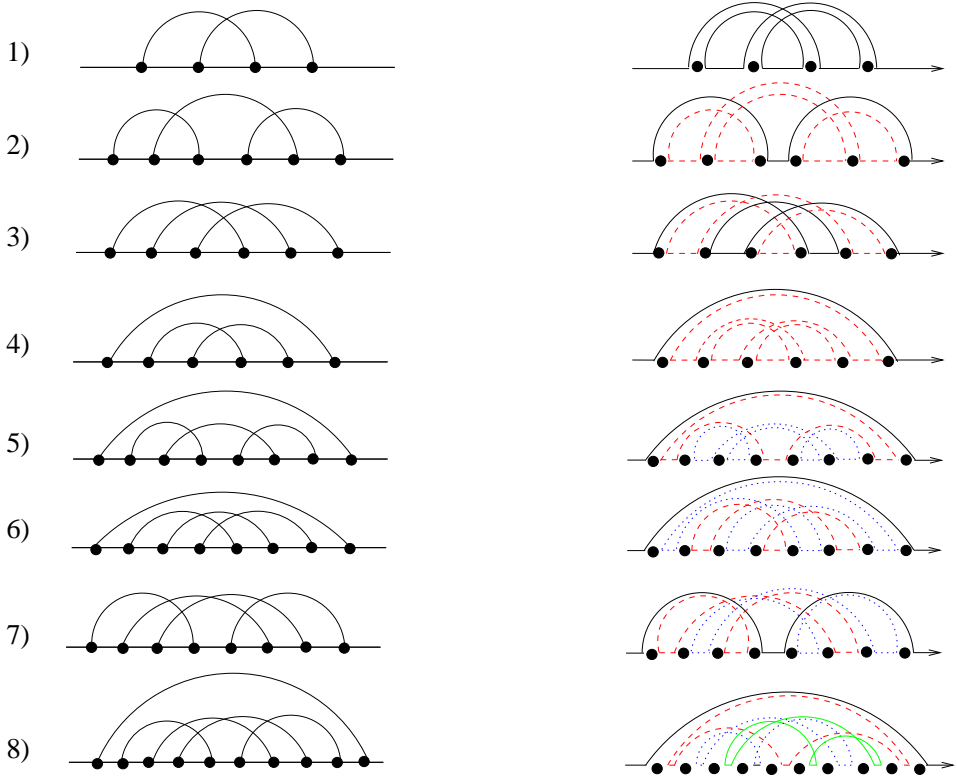,width=30pc}  
}  
\caption{List of all eight irreducible diagrams with genus $g=1$  
(from \cite{PTOZ}) and their representation with double line notation, on the left column and right column, respectively.}  
\label{eightplot}  
\end{figure}

\begin{figure}[-ht]  
\centerline{  
\epsfig{figure=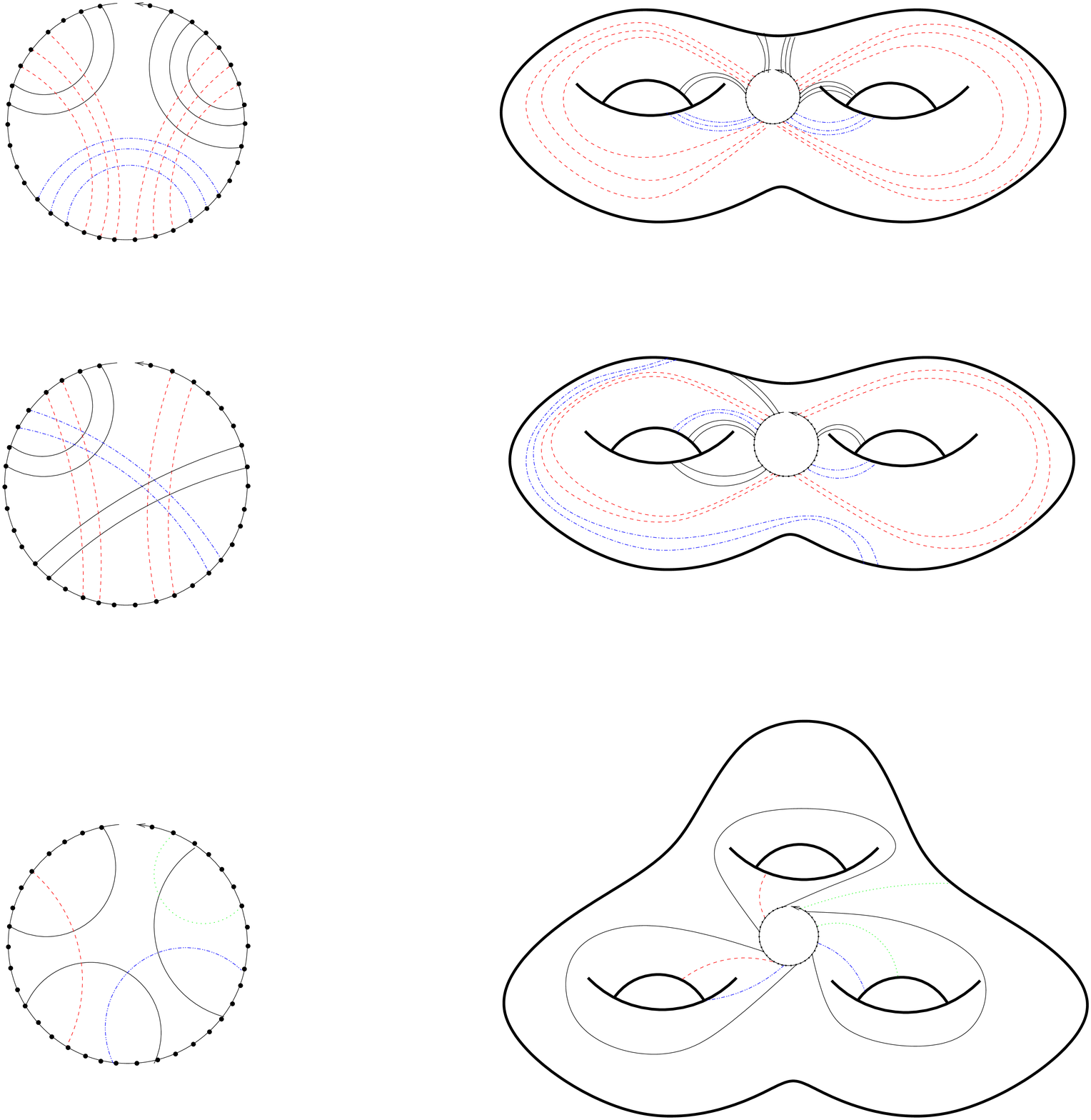,width=30pc}  
}  
\caption{Example of RNA pseudoknots with higher genus. The first two plots correspond to the diagrams a) and c) of Figure \ref{rivaseddyplot} with genus $g=2$. The third plot is an example with genus $g=3$.}  
\label{higherplot}  
\end{figure}

   
Let us discuss the case when there is a pseudoknot. For simplicity, we 
consider a ``kissing hairpin'' pseudoknot.  In this case the 
corresponding disk diagram is not planar, and has crossing arcs (see 
for instance Figure \ref{kisshairpinplot}). After drawing the disk 
diagram in double line notation, and counting the number of vertices, 
edges and faces, one gets that $\chi=-1$ this time. This value has a 
precise geometrical meaning.  In fact, the Euler characteristic of a 
surface (or of a manifold in general) is closely related to its {\it 
genus} $g$, i.e. the number of ``handles'' of the surface.  Namely if 
the manifold is orientable (as the disk diagram is, since the oriented 
circle line defines naturally an orientation of all the elements of 
the diagram), then one has $\chi = 2 - 2g-c$ where $c$ is the number 
of punctures ($c=1$ in the case we consider here, with only one RNA 
strand). It follows that a kissing hairpin is represented by a disk 
diagram with genus $g=1$. One concludes then, that such a disk diagram 
can be drawn on an oriented manifold with one handle, that is a {\it 
torus} (which is a doughnut-shaped surface formed by taking a cylinder 
and joining the two circular ends together, see Figure \ref{torusplot}).  This 
procedure can be extended easily to cases with more complex 
pseudoknots. For instance, the three diagrams of Figure 
\ref{rivaseddyplot} have genus $g=2$, $g=1$ and $g=2$, 
respectively. In Figure \ref{eightplot} there is a graphical 
representation of all 8 types of irreducible pseudoknots with genus 
$g=1$ (from \cite{POZ}) and in Figure \ref{higherplot} 
there is some examples of pseudoknots with a higher genus.\\ 
 
Thus we have a simple way to classify pseudoknots. This classification 
corresponds exactly to the series expansion of the partition function 
of the matrix model proposed in \cite{OZ}. There, the series is in 
powers of the form $N^{-2g}$, where $N$ is the size of the matrix, and 
$g$ is the genus of the corresponding set of diagrams. In the next 
section, we will exploit the same idea and show how one can 
control the topological character of pseudoknots in a statistical 
mechanical model for RNA secondary structures with pseudoknots.

\sect{Statistical mechanics model of RNA structures with pseudoknots}
\label{statsection} 
  
In almost all the energy models for RNA which have been proposed in  
recent years, the thermodynamical properties of a single stranded RNA  
are studied by means of a partition function of the form  
\begin{eqnarray}  
{\cal Z}_{RNA}&=&\int \prod_{k=1}^L \, d \mathbf  
{r}_k \,\, \sum_{C_{ij}} f(\{ \mathbf{r}\})   
e^{-\frac{1}{k_B T} U(C_{ij}, \{  \mathbf{r} \} )}  
\sim\sum_{C_{ij}} \omega(C) e^{-\frac{1}{k_B T} E(C) }= \nonumber \\  
&=&  
\sum_{C_{ij}} e^{-\frac{1}{k_B T}\left[ E(C)-T S(T,C) \right]} \, ,  
\label{ZRNA}  
\end{eqnarray}  
where $T$ is the temperature, $k_B$ is the Boltzmann constant,
$\mathbf{r}_k$ is the three-dimensional position vector of the $k$-th
nucleotide, $f(\{ \mathbf{r}\})$ takes into account the geometry and
the constraints of the chain of nucleotides, the function $U$ takes
into account the energetics coming from the pairing and stacking of
base pairs, and the sum over $C_{ij}$ is the sum over all possible
contact matrices for a given primary structure. The function
$\omega(C)$ is proportional to the number of configurations having the
same contact matrix $C$, and therefore its logarithm is just the
entropy factor associated to the polymeric nature of the
sugar-phosphate backbone. The free energy of a given configuration
${\cal F(C)} \equiv E(C)-T S(T,C)$ is the sum of several
contributions, both of energetic ( $E(C)$ ) and entropic nature (
$S(C)$ ): Watson-Crick and wobble base pairs, stacking energies,
terminal mismatches and dangling energies, special triloops and
tetraloops, entropy contributions (internal loops, bulges, hairpin
loops),
penalty factors for terminal-AU in helices, for asymmetries etc.  All
these terms have been determined empirically, and they are called
``Turner energy rules'' \cite{Tur}. For more details see
\cite{ZTM}. When pseudoknots are excluded, the sum in eq. (\ref{ZRNA}) is 
restricted over contact matrices that correspond to planar diagrams
only.  As we mentioned already in the introduction, the partition
function ${\cal Z}_{RNA}$ without pseudoknots can be calculated
efficiently by deterministic algorithms (dynamic programming)
\cite{caskill}: the most popular ones are perhaps the ``mfold
package'' by M.Zuker et al. \cite{z2003,mathews} and the ``RNA
Vienna package'' by I.Hofacker et al. \cite{ivovienna}\footnote{They
are available on-line at {\texttt
www.bioinfo.rpi.edu/applications/mfold/} and {\texttt
www.tbi.univie.ac.at/}, respectively.}.  When pseudoknots are
included, the sum in eq. (\ref{ZRNA}) is unrestricted and, as we
described in the previous Section, this leads to topology
fluctuations. This situation is very common also in other areas of
Physics (e.g. dynamical triangulations \cite{Mau}, random surfaces or
quantum gravity \cite{Amb}, quantum field theory \cite{thooft}), and
there are now standard ways to deal with it. The idea is to introduce
an additional parameter $\mu$, which is a topological ``chemical
potential'', and to consider the partition function:
\begin{equation}  
{\cal Z}_{RNA}(\mu)= \sum_{C_{ij}}   
e^{-\frac{1}{k_B T}\left[ E(C)-T S(T,C)+\mu g(C) \right]} \, , 
\label{ourZRNA}  
\end{equation}  
where $g(C)$ is the genus of the configuration associated to the contact 
matrix $C$. The ``chemical potential'' $\mu$ allows a simple control over the 
topological character of the pseudoknots in the statistical ensemble 
at thermal equilibrium. It is also directly related to $N$ (the size 
of the matrix) in the matrix model formulation of \cite{OZ}: 
\[  
{\cal Z}_{Matrix} \sim 1+\frac{Z_1}{N^2}+\frac{Z_2}{N^4}+\ldots \, ,
\]  
with $\mu=-2 k_B T \log(N)$.  The advantage here is that the energy  
function $E(C)$ can be more realistic than the one in \cite{OZ}.
  
The model without chemical potential, i.e. $\mu=0$, corresponds to the
case where there are no restrictions on the possible fluctuations of
the topology.  On the other hand when $\mu$ is very large, all the
configurations with $g>0$ are suppressed by the Boltzmann weight, and
in this case one recovers the planar limit (i.e. RNA secondary
structures without pseudoknots). One might expect then a phase
transition associated to the formation of pseudoknots. A natural order
parameter is the average genus of a RNA structure with pseudoknots
which can simply be recovered by taking the logarithmic derivative of
the partition function
\begin{equation}  
\langle g(\mu) \rangle  
=   
-k_B T  \frac{\partial}{\partial \mu} \log {\cal Z}_{RNA}(\mu) \, . 
\end{equation}  
To our knowledge there are no available experimental data about 
the dependence of the genus of RNA molecules on the 
temperature. Informations and inputs from experiments would be highly 
desirable. 
  
\begin{figure}[-ht]  
\centerline{  
\epsfig{figure=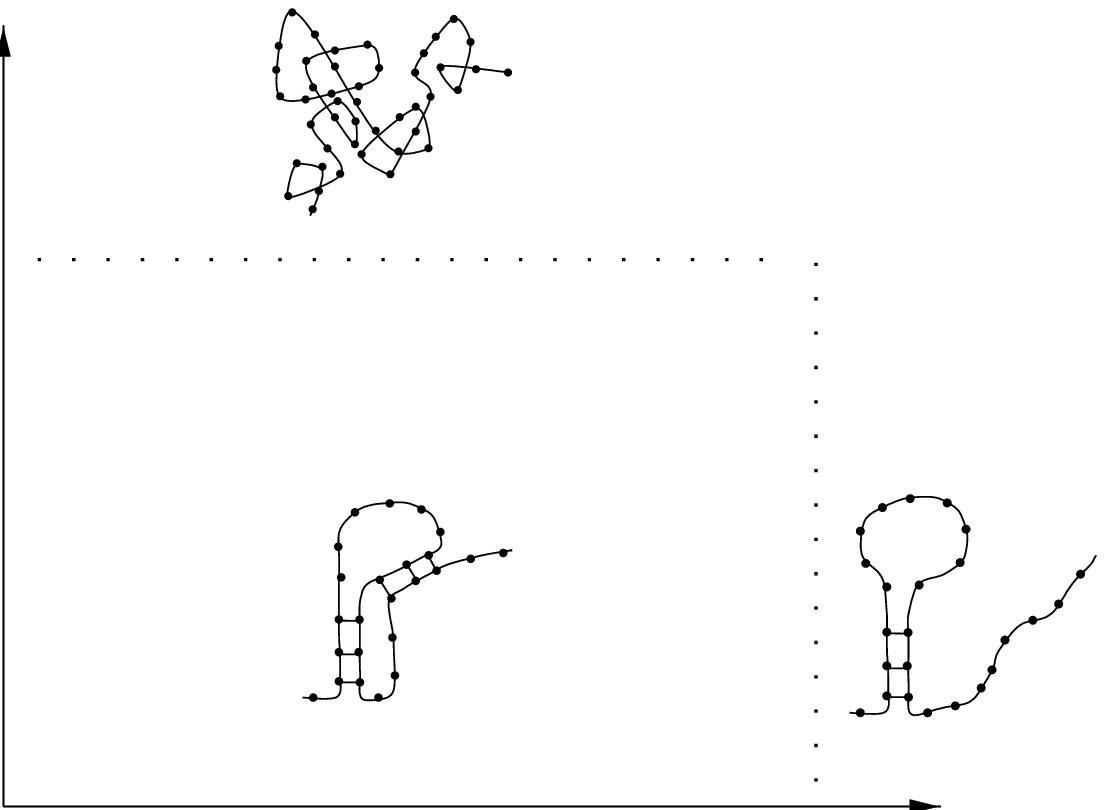,width=15pc}  
\put(-190,120){$T$}  
\put(-35,-9) {$\mu$}  
\put(-95,120) {denaturated phase}  
\put(-5,50) {planar limit}  
\put(-170,10) {phase with pseudoknots}  
}  
\caption{Qualitative structure of the phase diagram in the  $\{\mu,T\}$ plane.}  
\label{phaseplot}  
\end{figure}

Figure \ref{phaseplot} displays the expected phase diagram in the
plane $\{\mu,T\}$ of our model. At high temperature, the RNA is always
in a fully denaturated phase. At lower temperature and large $\mu$ the
secondary structures without pseudoknots are the dominating
configurations.  The interesting part of the diagram is for lower
values of $\mu$, where possibly $ \langle g(\mu) \rangle \; \neq 0$
and pseudoknots are present.
 
Even if eq. (\ref{ourZRNA}) can in principle deal with pseudoknotted
RNA molecules, it is fair to say that for any realistic energy
function, the model is rather unlikely amenable to an analytic
solution. Moreover any dynamic programming approach has been shown to
be computationally very demanding even for pseudoknots with genus
$g=1$ \cite{PTOZ}. Hence a stochastic algorithm for studying the model
eq.  (\ref{ourZRNA}) is probably the only feasible way. In the next
Section we describe in details a Monte Carlo algorithm for the
simulation of the model of eq. (\ref{ourZRNA}).
  
\sect{A Monte Carlo algorithm for RNA pseudoknots prediction} 
\label{montesection}   
  
  
A well-known method for generating a set of configurations which are
distributed according to a given Boltzmann weight is the Monte Carlo
method.  It is a standard method of modern computational analysis and
we refer to \cite{MC,frenkel} for a review and an introduction on the
subject.  In recent years, it has been also used for the prediction of
RNA secondary structures in various contexts. In particular, our
proposal can be thought of as a generalization of the Monte Carlo
method described in \cite{Monte} where the authors considered only RNA
secondary structures without pseudoknots. We aim to apply this method
to the statistical ensemble defined by eq. (\ref{ourZRNA}).
  
The sum over all the RNA configurations in eq. (\ref{ourZRNA}) 
contains many terms. In general, the total number of RNA 
configurations (planar and non planar configurations) grows like $L!$ 
for a sequence of length $L$.  The number of RNA configurations with a 
fixed genus $g$ grows exponentially with $L$: a detailed analysis 
for the number of diagrams with genus $g=0$ (i.e. planar diagrams) can 
be found in \cite{combinatorics}.\footnote{An analysis for structure 
with higher genus similar to the one in \cite{combinatorics} is still 
lacking. We expect that the matrix field theory model introduced in 
\cite{OZ} can shed some light on this issue.} 
Since the number of secondary structures on a surface with fixed genus
grows exponentially, one expects that a brute force Monte Carlo
importance sampling would be rather ineffective for a not too-short
RNA sequence, in a reasonable amount of computation time.  For that
reason we decided to use the standard Metropolis method \cite{Metro}. The
Metropolis method is an efficient and simple scheme for generating a
set of configurations distributed according to a given probability
function, by means of a random walk in the configuration space. In our
case, the Metropolis Monte Carlo method generates a set of $n$ RNA
configurations $\{C^{(0)},C^{(1)},\ldots,C^{(n)}\}$, such that
$\lim_{n
\to \infty} n_{C}/n=P(C)$, where $P(C)$ is the given  
probability distribution (e.g., the Boltzmann distribution $P(C) =
{\cal Z}^{-1} \exp[\left( E-T S+\mu g \right) / k_B T ]$ and $n_C$ is
the number of configurations of type $C$ in the statistical
ensemble. Each element $C^{(k)}$ of the sequence is generated by
accepting or rejecting a random configuration. In the following we
give a complete description of the Metropolis Monte Carlo algorithm
for RNA pseudoknots predictions:
  
\begin{itemize}  
  
\item Step 1: Pick an initial configuration $C^{(0)}$:  
A simple initial configuration can be the fully denaturated state of 
RNA, i.e.  the contact matrix is the matrix with all zero entries and 
the respective permutation involution is the identity permutation 
$\sigma=\left( 
\begin{array}{cccc} 
1&2&\ldots&L\\ 
1&2&\ldots&L 
\end{array} 
\right)$. Set the variable $n=1$.

\item Step 2: Pick a trial configuration $C^{(n)}$ (by deforming   
the configuration $C^{(n-1)}$). Such an operation is called ``Monte  
Carlo move'' $C^{(n-1)} \to C^{(n)}$. Compute the probability ratio  
\begin{equation} 
\rho=\frac{P(C^{(n)})}{P(C^{(n-1)})} \, . 
\label{rho} 
\end{equation} 
Pick a random number $x$ with value between 0 and 1.  If $x \leq \rho$ 
accept the configuration $C^{(n)}$ as the new configuration. 
Otherwise refuse it and keep $C^{(n-1)}$ as new configuration, 
i.e. put $C^{(n)}=C^{(n-1)}$ . Increase the variable $n$ by one.

\item  Step 3: repeat Step 2 for $n_{max}$ times, where $n_{max}$ is a  
sufficiently large number.   
  
\end{itemize}

The most relevant aspect of this method is that, at large $n_{max}$, 
one can generate an ensemble of configurations with the probability 
distribution $P(C) = {\cal Z}^{-1} \exp[-(E(C)-T S(T,C) +\mu 
g(C))/(k_B T)]$, simply by computing probability ratios.  Therefore, 
this method is extremely useful as it avoids the need of computing the 
partition function ${\cal Z}$ of the system, a computational task that would be 
surely intractable for long RNA sequences. 
\begin{figure}[-ht]  
\centerline{  
\epsfig{figure=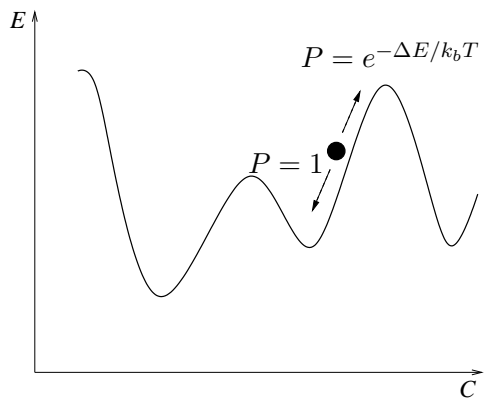,width=15pc}  
\put(-90,85) {$P=1$} 
\put(-70,125) {$P=e^{-\Delta E/k_b T}$} 
}  
\caption{The Metropolis algorithm accepts a configuration with lower energy
with probability $P=1$. It can also accept a configuration with higher
energy, with probability $P=e^{-\Delta E/K_b T}$, where $\Delta E$ is
the energy difference.}
\label{energyplot}  
\end{figure}

\subsection{Configurational changes (Monte Carlo moves)}  
      
At large $n_{max}$, the above algorithm is guaranteed to generate a 
set of configurations with the probability distribution $P(C)$, under 
few assumptions. Two essential requirements are that the 
Monte Carlo moves have to be {\it ergodic} and satisfy the so-called 
{\it detailed balance condition}. Ergodicity essentially means that every 
point in the configuration space can be reached in a finite number of 
Monte Carlo steps from any other point. The detailed balance condition 
here simply means that the Monte Carlo moves are symmetric, i.e.  the 
probability of proposing a Monte Carlo move $C \to C'$ is the same as 
of proposing the move $C' \to C$. 
 
\begin{figure}[-ht]  
\centerline{  
\epsfig{figure=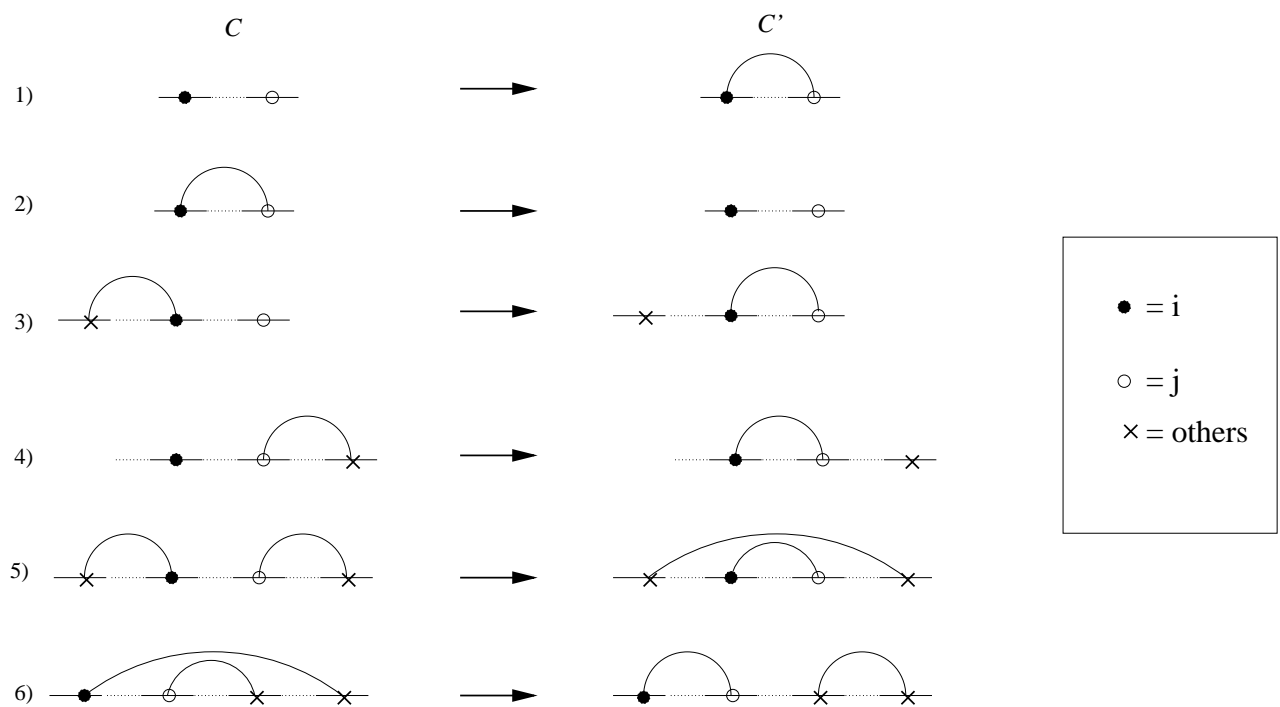,width=40pc}  
} 
\caption{Monte Carlo moves for an allowed base pair $i-j$ of a  
RNA secondary structures with pseudoknots. The move $1)$ adds a
base pair. The move $2)$ removes a base pair. The moves $3)$ and $4)$
shift a base pair.  The move $5)$ and $6)$ swap two base-pairs, when
possible.}
\label{MCmovesplot}  
\end{figure}

\begin{figure}[-ht]  
\centerline{  
\epsfig{figure=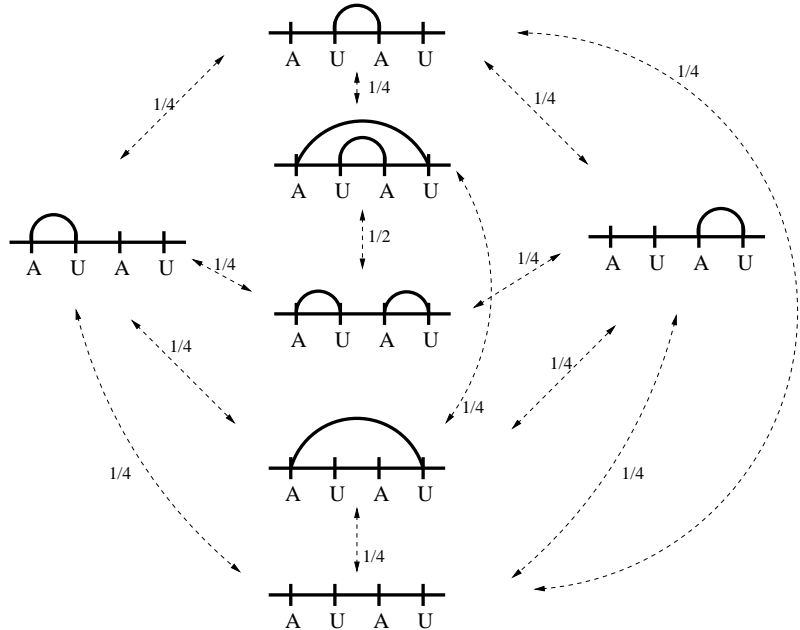,width=25pc}  
}  
\caption{Space of the configurations for the sequence $\{A,U,A,U\}$. The 
arrows indicate the Monte Carlo moves and their probability rate.} 
\label{confplot}  
\end{figure}

We describe now the Monte Carlo moves for  
RNA folding. First, at the beginning of the 
simulation, it is useful to make some book-keeping by storing in the memory the list of 
all the allowed base-pairs (i.e. that are only of the type 
A$\bullet$U, C$\bullet$G or G$\bullet$U).  Such an information can be 
stored in $L$ vectors $l_i$, $i=1,\ldots,L$, as follows: the 
nucleotide in position $i$ can be paired to $n_i$ 
possible other nucleotides, namely with the ones in position $l_i(1), 
l_i(2),\ldots,l_i(n_i)$ and nothing else. For example, if the primary 
structure is $\{AGCU\}$ then we have: 
\begin{equation} 
l_1=[4] \, , \quad l_2=[3,4] \, , \quad l_3=[2] \, , \quad l_4=[1,2]   
\, . 
\end{equation} 
The creation of such a list of possible base-pairs does not slow down
the total algorithm since it is an $O(L^2)$ operation which is done
only once. Now we want to extract one element from the list of $L$
vectors with uniform probability. This can be done as follows. Let $T
\equiv \sum_h n_h$, and let pick up a uniform integer random number $\tau$
between $1 \leq \tau \leq T$. Then take the highest integer number $i$
such that $\sum_{h=1}^{i} n_h \leq \tau$, and define
$y\equiv\tau-\sum_{h=1}^{i} n_h+1$. Obviously $1 \leq y \leq T$ holds
true. Consider the pair of bases $i$ and $j \equiv l_i(y)$.  The
base-pair $i-j$ has been extracted randomly with uniform probability
in the set of all possible base-pairs, for the given RNA sequence.  The
Monte Carlo move $C\to C'$ is then generated as follows:
\begin{itemize}  
  
\item If the configuration $C$ is such that both the base in $i$ and in $j$ are
free (i.e. $\sigma_{C}(i)=i$ and $\sigma_{C}(j)=j$) then add the link
$i-j$ (i.e. put $\sigma_{C'}(i)=j$ and $\sigma_{C'}(j)=i$). We call
this Monte Carlo move ``add a base pair'' (see case 1 of figure
\ref{MCmovesplot}).

\item If the configuration $C$ is such that there is arc between $i$ and $j$
(i.e. $\sigma_{C}(i)=j$) then remove the link $i-j$ (i.e. put
$\sigma_{C'}(i)=i$ and $\sigma_{C'}(j)=j$). We call this Monte Carlo
move ``remove a base pair'' (see case 2 of figure \ref{MCmovesplot}).

\item If the configuration $C$ is such that either the base in $i$ or the
base in $j$ is linked to some other base, (i.e. $\sigma_{C}(i)=i$ and
$\sigma_{C}(j) \neq j$, or $\sigma_{C}(j)=j$ and $\sigma_{C}(i) \neq i$)
then move the link back to $i-j$, by overriding any former link
(i.e. put $\sigma_{C'}(i)=j$ and $\sigma_{C'}(j)=i$).  We call this
Monte Carlo move ``shift a base pair'' (see case 3 and 4 of figure
\ref{MCmovesplot}).

\item If the configuration $C$ is such that the base $i$ is linked to an other
base $k_1$ and $j$ is linked to an other base $k_2$ and the base-pair
$k_1-k_2$ is possible, then swap the links (i.e. put
$\sigma_{C'}(i)=j$, $\sigma_{C'}(j)=i$, $\sigma_{C'}(k_1)=k_2$,
$\sigma_{C'}(k_2)=k_1$).  We call this Monte Carlo move ``swap a base
pair'' (see case 5 and 6 of figure
\ref{MCmovesplot}).

\item If none of the above cases applies then do not update the
configuration, i.e. put $C'=C$.

\end{itemize} 
See Figure \ref{MCmovesplot} for a summary of these Monte Carlo moves, and Figure \ref{confplot} for a simple example.\\

These Monte Carlo moves obviously satisfy the detailed balance 
condition. In fact the probability of creating a link between $i$ and 
$j$ when $i$ or $j$ are link-free (or at least one of them), or of 
removing the link when they are already linked is always $P_{ij}=2/T$, 
and thus it is symmetric. In the case where $i$ and $j$ are already 
linked to different bases $\sigma(i)$ and $\sigma(j)$, then a link is 
put between $i$ and $j$ only if $\sigma(i)$ can be connected to 
$\sigma(j)$ as well. In this case the reverse move also occurs with 
the same probability rate, thus it is a symmetric move. Moreover, the set of 
Monte Carlo moves are ergodic. The key 
observation is that such moves correspond to transpositions in the 
space of permutation involutions.  Since any configuration of RNA 
secondary structure with pseudoknots can be uniquely represented by a 
permutation involution, and since any permutation can be obtained by a 
suitable finite sequence of transpositions 
\cite{knuth3}, it follows that any RNA secondary structure with pseudoknots can be  
generated with a finite number of the Monte Carlo moves described 
above. 
 
Few comments are in order.  First of all, other sets of Monte Carlo
moves are possible of course. Several authors introduced collective
moves, where several links are updated at the same time (as opposed to
one by one as we propose).  The advantages are a general speed-up of
the computing time, and a more effective simulation as far as overcoming
the energy barriers.  In the present work, we prefer to keep the code
as simple as possible by using a set of ``local'' moves, and to focus
on testing its effectiveness when dealing with RNA pseudoknots.
Second, both the generation of the Monte Carlo moves and the
Metropolis method require a good (pseudo)random number generator, in
order to avoid biases in the output which may be very difficult to
detect.  For a good introduction to random number generators we refer
the reader to
\cite{knuth2} and \cite{NumRecipes}.  
Finally, as in all stochastic algorithms, one has to be able to 
estimate the statistical errors of the Monte Carlo prediction.  As 
this method generates an ensemble of configurations 
$\{C^{(0)},C^{(1)},\ldots, C^{(n_{max})}\}$, distributed according to 
the probability distribution $P(C)$, then one can compute ensemble 
averages of any quantity ${\cal A}(C)$ simply by: 
\begin{equation}  
\langle {\cal A} \rangle =\frac{1}{n_{max}}\sum_{i=1}^{n_{max}} {\cal A}(C^{(i)})  \, . 
\end{equation}  
  
The error associated to this observable scales like $1/\sqrt{N}$ where 
$N$ is the number of independent measurements. It is important to 
note that $N$ is not usually equal to $n_{max}$ since in general the 
configurations generated by any Monte Carlo algorithm are 
correlated. One can deal with this issue in two ways.  One can compute 
the autocorrelation length $\xi$ of the sequence of the RNA 
configurations generated by the Monte Carlo algorithm and then 
subsample the same set of configurations, keeping one configuration 
every $\xi$ and skipping all the configurations in between 
\cite{Geyer92}.  An other possibility is to keep all the 
configurations of the sequence, and compute the statistical error by
taking into account the existence of correlations (The error is
usually bigger than the simple standard deviation of the data). A
well-known technique for computing the statistical error of a set of
correlated data is the so-called {\it jackknife method}. For an
introduction to this method, we refer the reader to
\cite{jackknife}. There are also other techniques which can be found in
\cite{Liu}.
  
\subsection{Energy update}  
According to the Metropolis method, the ratio $\rho$ (Step 2 of the
algorithm, eq. (\ref{rho})) is given by
\begin{equation}  
\rho=\exp\left[-\frac{1}{k_B T}(\Delta E-T \Delta S+\mu \Delta g )\right] \, , 
\end{equation}  
where 
\begin{eqnarray} 
\Delta E&=&E(C^{(n)})-E(C^{(n-1)}) \nonumber \, ,\\ 
\Delta S&=&S(C^{(n)})-S(C^{(n-1)}) \nonumber \, , \\ 
\Delta g&=&g(C^{(n)})-g(C^{(n-1)}) \, . \nonumber 
\end{eqnarray} 
Since the Monte Carlo moves are local (i.e. they involve only a small 
part of the RNA sequence) the computation of $\Delta E$, $\Delta S$ 
and $\Delta g$ is usually easier and faster than computing the full 
functions $E(C)$, $S(C)$ and $g(C)$. 
  
We consider first $\Delta g$, and we provide an efficient algorithm 
for computing it.  According to eq. (\ref{genus}), the genus is given 
by $g=(1-\#V+\#E-\#F)/2= 
(1-L+(L+n_{arcs})-n_{loops})/2$.  
Therefore:  
\begin{equation}  
\Delta g=\frac{1+\Delta n_{arcs}-\Delta n_{loops}}{2} \, ,  
\end{equation}  
where 
\begin{equation} 
\Delta n_{arcs}= 
\left\{ 
\begin{array}{ll} 
-1& \mbox{ for a ``remove the base-pair $i-j$" move,}\\ 
1 & \mbox{ for a  ``add the base pair $i-j$" move,}\\ 
0 & \mbox{ for a ``shift or swap the base pair $i-j$" move.} 
\end{array} 
\right.   
\end{equation} 
The difference $\Delta n_{loops}$ can be computed by considering the
loops containing the bases $i$ and $j$. In principle there are 4
possible independent loops, two about $i$ and two about $j$ (see
Figure \ref{loopsplot}). The connectivity of the RNA molecule can be
such that the four loops are not independent. In Appendix
\ref{appendix} we describe an algorithm for computing the actual number of
independent loops. Then it is sufficient to run the algorithm over
$C^{(n)}$ and $C^{(n-1)}$ and compute the difference of loops, that is
$\Delta n_{loops}$.
\begin{figure}[-ht]  
\centerline{  
\epsfig{figure=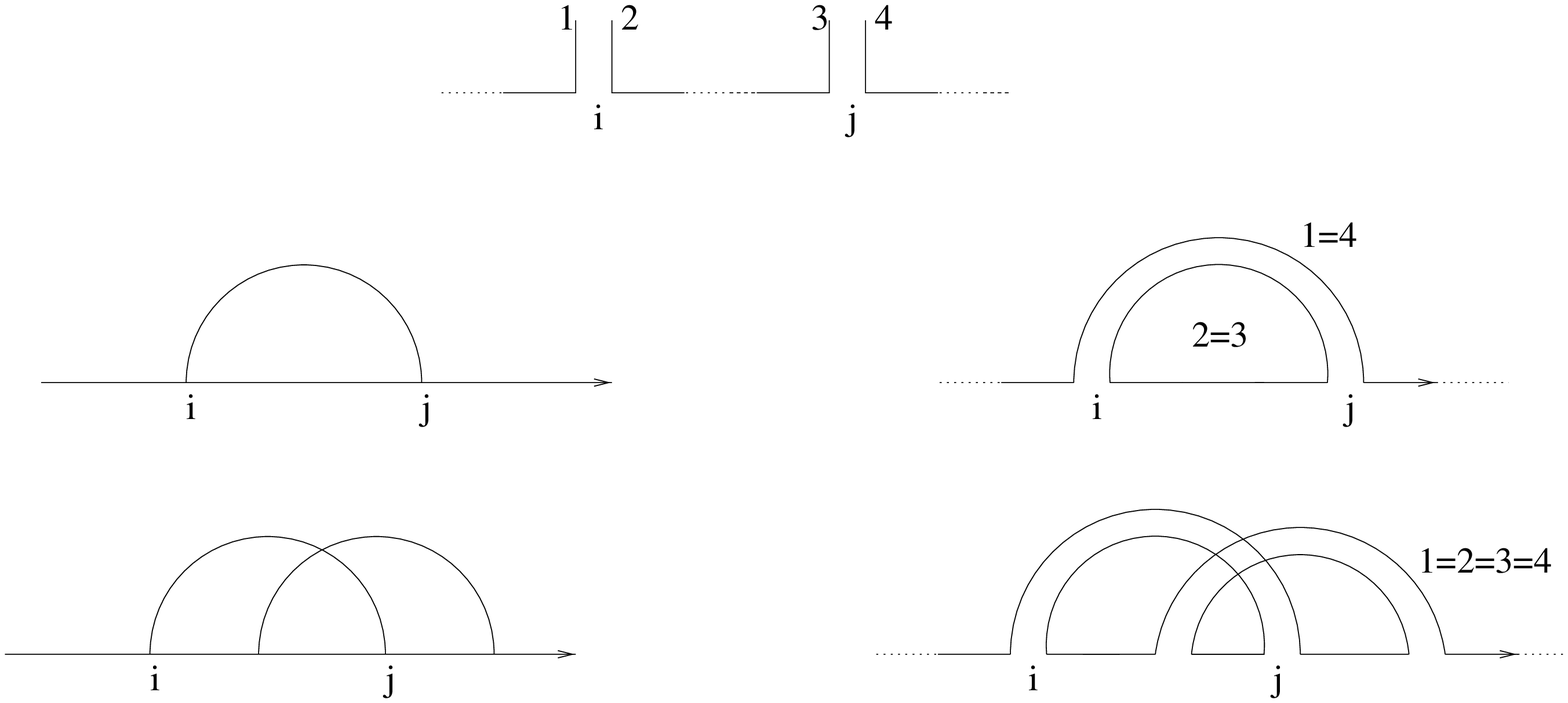,width=20pc}  
}  
\caption{Two given bases $i$ and $j$ usually belong to $4$ loops, when drawing the arc with the double-line notation. The loops 
are not always independent. Here there are two examples: a case with
$2$ independent loops (top), and a case with only one
independent loop (bottom).}
\label{loopsplot}  
\end{figure}  
  
Secondly, the calculation of $\Delta E$ is more problematic. There is
not yet an RNA energy model for any given topology. The most studied
and well-defined model both from a theoretical and experimental point
of view is for the spherical topology (i.e. genus$=0$, that is RNA
secondary structure without pseudoknots)\cite{ZukEn}. The set of
empirical ``Turner'' energy rules can be generalized in order to
describe some simple class of pseudoknots \cite{gultyaev} but the
general case for {\it any} topology is still lacking.  For $\Delta S$
the situation is slightly better, as one can model the configurational
entropy of the RNA structure by using the theory of polymers (as
already presented in \cite{isa1} or \cite{gultyaev}) and the inclusion
of pseudoknots is in principle feasible. Therefore, in our
numerical simulation we will use the ``Turner'' energy model, and even
if this is not quite appropriate for higher topology, we expect the
corrections to be small with respect to the over-all energy scale. We
remind the reader that the purpose of this paper is to propose a new
approach for the study of pseudoknots formation in RNA secondary
structure. Thus at this point, it is reasonable to perform a
preliminary analysis based on an approximate energy model.
When a more complete energy model (including all the 
topologies) will be available, it will be sufficient to include it in the 
calculation of $\Delta E$ in our algorithm. 
 
\subsection{The ``simulated annealing" method}  
One of the major problems about the Monte Carlo simulation of RNA
folding, is that the energy landscape is usually very rough with
metastable valleys separated by energy barriers which are ``high''
compared to the energy involved in each Monte Carlo move. This is a
general situation in thermodynamic systems with many degrees of
freedom (e.g.  glasses, polymers, proteins etc.).  where in addition
to the global minimum energy configuration there may be many local
minima separated by high energy barriers. The worst consequence is
that the system can be trapped for a long time in such local minima
and the Monte Carlo exploration of the energy landscape is no longer
effective. In order to over come this problem with RNA M.Schmitz and
G. Steger in \cite{Monte} proposed the use of a computational
technique called ``simulated annealing'' method. It is a classical
method which has been introduce for finding the minimum energy
configuration of a system with a very rough energy landscape
\cite{kirk}. We briefly describe the algorithm: 
 
\begin{itemize} 
\item Step 1: generalize the partition function eq. (\ref{ourZRNA}) to the form: 
\begin{equation} 
{\cal Z}_{RNA}= \sum_{C_{ij}} e^{-\frac{1}{k_B \Theta}\left[ E(C)-T 
S(T,C)+\mu g(C) \right]} \, , 
\label{ourZRNATheta} 
\end{equation} 
and initialize $\Theta=\Theta_{max}>T$. 
\item Step 2: Starting from an initial configuration $C^{(0)}$ (e.g., 
the fully denaturated RNA configuration) sample $n$ configurations by 
means of the Metropolis Monte Carlo method applied to 
eq. (\ref{ourZRNATheta}). 
\item Step 3: Go to Step 2, and replace $\Theta$ by a lower value, and 
$C^{(0)}$ by $C^{(n)}$. Repeat this step until the temperature of the 
system is equal to $T$. During the Monte Carlo process keep track all 
the time of the configuration with the lowest energy. 
\end{itemize} 
One can show that usually, the global minimum can be obtained by using
a logarithmic rate \cite{annea}.  In practice, other annealing
schedules are possible: linear, hyperbolic, exponential, power-law
schedules are often implemented.

Assuming that at low temperature an RNA molecule assumes a
configuration which corresponds to the minimum energy, we can find
such a configuration by using the simulated annealing method, starting
the simulation with a value of $\Theta$ well above the melting
temperature (say a few hundred degrees Celsius).  A first check of
this method is whether we can reproduce the results produced by
deterministic algorithms such as ``mfold'' \cite{z2003} or the
``Vienna Package'' \cite{ivovienna}. For that purpose, it is
sufficient to use the ``Turner'' energy model and run our algorithm
with a large value of the chemical potential $\mu$. Our preliminary
tests showed that the minimum can be easily found for sequences with
length up to around 300 bases. For longer RNA sequences, the
simulation time increases and the minimum may be harder to find. In
this cases we can use an additional feature of our model. In fact our
approach offers also the interesting possibility of using the chemical
potential for overcoming the energy barriers. It means that we can
apply a ``simulated annealing'' method on $\mu$ rather than on $T$.
Thus, starting with a low value of $\mu$ (where all the topologies
with any genus are allowed) the Monte Carlo simulation can quickly
explore regions which are very distant from each other in the energy
landscape. Then by slowly increasing the value of $\mu$ we gradually
constrain the simulation to select only planar configurations
(i.e. secondary structures without pseudoknots), and the minimum
energy configuration eventually.  During this process, that is for
intermediate values of $\mu$, many configurations in thermal
equilibrium are generated, and in general they correspond to diagrams
with $\langle g \rangle \neq0$, i.e. RNA configurations with
pseudoknots. These configurations are the prediction of our algorithm
and they should be compared with the experimental data. It is at this
level that the value of $\mu$ can be tuned, in order to fit the
data. The results of our preliminary investigations in this region of
the phase diagram are very encouraging and promising. However, in this
paper we limit ourselves to the description of this new method and of
our algorithm. The results of our simulation will be published
shortly.

\sect{Conclusions}\label{conclsection}

In this paper we propose a new approach to the problem of RNA folding
with pseudoknots. We start from a classifications of RNA pseudoknots
based on their graphical representation by means of disk diagrams.  A
generic disk diagram is usually not planar, i.e. cannot be drawn on a
plane surface without crossing lines. However, if the surface has a
high enough genus (i.e. a sufficient number of ``handles''), the
diagram can always be drawn on that surface without any crossing.  The
precise correspondence is obtained by using a famous theorem by Euler,
and it precisely corresponds to the topological classifications of RNA
pseudoknots already introduced in \cite{OZ}. Then we propose a
statistical mechanics model where the formation of RNA pseudoknots is
associated with fluctuations of the topology (eq. (\ref{ourZRNA})). In
order to do that we introduce a parameter, the topological ``chemical
potential'', which controls the rate of pseudoknots formation, and can
be obtained by fitting experimental data. We then discuss the
qualitative structure of the phase diagram for the RNA molecule in the
plane $\{\mu,T\}$ and its interpretation. Finally we describe a Monte
Carlo algorithm for the prediction of RNA pseudoknots. It is based on
a standard Metropolis algorithm coupled to the ``simulated annealing''
method and we provide an explicit description of its implementation
and use.  A numerical investigation of this technique and the phase
diagram is under way and will be published shortly.

\indent  
 
\noindent  
\underline{Acknowledgments}: 
We wish to thank the organizers of the EUROGRID meeting on ``Random
geometry: theory and applications'' at Les Houches (France) on March
2004, where this work has been first presented. GV acknowledges the
support of the European Fellowship MEIF-CT-2003-501547.

\begin{appendix}
   
\sect{An algorithm for computing $n_{loops}$} 
\label{appendix}
 
In this Appendix we describe an algorithm for computing the number of
independent loops adjacent to the $i$-th and $j$-th nucleotides. It is
useful for computing the variation of the genus when one of the Monte
Carlo moves we described in Section \ref{montesection}.  The algorithm
we propose for counting the number of independent loops is based on
tracking a walk along the diagram starting from the base $i$ and
marking the loop with an identifying number (or color). Namely, we
represent the configuration $C$ by means of the permutation involution
$\sigma_{C}$ (as described in Section \ref{representsection}), and the
algorithm is:
 
\begin{verbatim} 
START  
v(1)=v(2)=v(3)=v(4)=0           % set the four flags to zero 
pos=i                           % the start position is i-th base 
color=1                         % using color 1 
v(1)=color                      % mark the first flag with the color in use 
do{                             % start the first loop 
   pos=sigma(pos)               % follow the permutation involution 
   if pos==i then v(2)=color    % check if it is either in i or j        
   if pos==j then v(4)=color             
   pos=shift(pos)               % shift move (along the RNA circle) 
   if pos==j then v(3)=color    % check if it is in i again 
  } while (position!=i)         % repeat until it returns at the starting point 
 
if v(2)==0 then{                % check if the second loop has been marked already 
  color=color+1                 % if yes, change color 
  pos=i                         % start again from i-th base 
  v(2)=color                    % mark the second flag  
  do{                           % repeat all the above for the second loop (at i) 
     pos=shift(pos) 
     if pos==j then v(3)=color 
     pos=sigma(pos) 
     if pos==j then v(4)=color      
     } while (pos!=i) 
} 
 
 
if v(3)==0 then{                % repeat all the above for the third loop (at j) 
   color=color+1 
   pos=j 
   v(3)=color 
   do{  
      pos=sigma(pos) 
      if pos==j then v(4)=color 
      pos=shift(pos) 
      } while (pos!=j) 
} 
 
if v(4)==0 then{                % the fourth loop at j is independent from the
   color=color+1                % previous ones if and only if it has not been
}                               % marked yet
 
nloops=color                    % the number of independent loops is the number 
                                % of used colors 
END  
\end{verbatim} 
$L$ is the length of the RNA sequence and the function {\texttt
shift(pos)} is just the increment by 1 of the variable {\texttt pos}
with period $L$, i.e. {\texttt shift(pos)=remainder(pos,L)+1}. At the
end of the algorithm the variable ``color'' contains the number of
independent loops $n_{loops}$. The algorithm runs in a time
proportional to $O(L)$.
\end{appendix} 
  
   

\end{document}